\documentclass[sigconf]{acmart}

\usepackage{array}
\usepackage[normalem]{ulem}
\usepackage{subcaption}
\usepackage{soul}
\usepackage{enumitem}
\usepackage{rotating}

\AtBeginDocument{%
  \providecommand\BibTeX{{%
    \normalfont B\kern-0.5em{\scshape i\kern-0.25em b}\kern-0.8em\TeX}}}
\copyrightyear{2024}
\acmYear{2024}
\setcopyright{acmlicensed}\acmConference[SIGIR '24]{Proceedings of the 47th International ACM SIGIR Conference on Research and Development in Information Retrieval}{July 14--18, 2024}{Washington, DC, USA}
\acmBooktitle{Proceedings of the 47th International ACM SIGIR Conference on Research and Development in Information Retrieval (SIGIR '24), July 14--18, 2024, Washington, DC, USA}
\acmDOI{10.1145/3626772.3657933}
\acmISBN{979-8-4007-0431-4/24/07}

\usepackage{booktabs} 
\usepackage{color}
\usepackage{arydshln}
\usepackage{multirow}
\usepackage{graphbox}

\begin{document}

\title{A Surprisingly Simple yet Effective Multi-Query Rewriting Method for Conversational Passage Retrieval}

\author{Ivica Kostric}
\orcid{0000-0002-0859-9762}
\affiliation{%
  \institution{University of Stavanger}
  \city{Stavanger}
  \country{Norway}
}
\email{ivica.kostric@uis.no}

\author{Krisztian Balog}
\orcid{0000-0003-2762-721X}
\affiliation{%
  \institution{University of Stavanger}
  \city{Stavanger}
  \country{Norway}
}
\email{krisztian.balog@uis.no}

\begin{abstract}
Conversational passage retrieval is challenging as it often requires the resolution of references to previous utterances and needs to deal with the complexities of natural language, such as coreference and ellipsis.  To address these challenges, pre-trained sequence-to-sequence neural query rewriters are commonly used to generate a single de-contextualized query based on conversation history.  Previous research shows that combining multiple query rewrites for the same user utterance has a positive effect on retrieval performance.  We propose the use of a neural query rewriter to generate multiple queries and show how to integrate those queries in the passage retrieval pipeline efficiently.  The main strength of our approach lies in its simplicity: it leverages how the beam search algorithm works and can produce multiple query rewrites at no additional cost.  Our contributions further include devising ways to utilize multi-query rewrites in both sparse and dense first-pass retrieval. We demonstrate that applying our approach on top of a standard passage retrieval pipeline delivers state-of-the-art performance without sacrificing efficiency.
\end{abstract}




\maketitle

\section{Introduction}
 
The main objective of a conversational search system is to effectively retrieve relevant answers to a wide range of information needs expressed in natural language~\citep{Anand:2021:SIGIR}.
A major difficulty lies in the conversational nature of the task, namely, that queries are often not standalone and need to be interpreted in the context of the user's previous queries as well as the system's answers to those~\citep{Zamani:2023:INR}.  
Commonly, \emph{query rewriting} (QR) addresses this by employing neural generative models to produce a single de-contextualized query at each conversation turn~\citep{Elgohary:2019:EMNLP-IJCNLP, Lin:2020:arXiv, Yu:2020:SIGIR}, and then feed that query to a retrieval pipeline.  While this works well in many cases, the rewritten query may incorrectly capture the underlying intent, which leads to the retrieval of non-relevant answers. The challenge arises from the discrete generation process, which does not accurately capture the underlying probabilities or importance of terms.

In this paper, we seek to improve retrieval performance by generating multiple queries and modeling the importance of terms based on their presence across the queries.  We leverage the beam search algorithm, commonly used in neural QR~\citep{Gao:2022:arXiv, Lin:2021:ACM}.  Instead of keeping track of only the highest-likelihood sequence in a greedy fashion, the algorithm tracks and considers the best $k$ sequences at each generation step.  We utilize the fact that the token probabilities are already computed in order to produce multiple rewrites at no additional cost.
Thus, the only modification we need to make to the original beam search algorithm is to return all tracked sequences and their associated probabilities, as opposed to the single most probable sequence.
The elegance of this method lies in its simplicity; it is computationally inexpensive yet remarkably effective.

The main research question driving our investigation is: \emph{How can we effectively and efficiently utilize multiple query rewrites in conversational passage retrieval?}  To answer this question, we take into consideration that retrieval can be performed using either sparse or dense retrieval methods. 
Sparse retrieval typically employs pseudo-relevance feedback techniques to expand the query and bridge the vocabulary gap.  Our method effectively performs both term-importance estimation and query expansion to represent the underlying information need better and improves MRR by 1.06--6.31 percentage points compared to using a single query rewrite. Dense retrieval, based on contextual neural language models, works better with natural language queries (in contrast to bag-of-words models of sparse retrieval).
However, it is computationally expensive and would scale linearly with the number of rewrites, rendering it impractical.  Instead, we represent all query rewrites jointly by merging them into a single vector representation in the learned embedding space by weighted average pooling. Our method outperforms a single-query retrieval by 3.52--4.45 percentage points in absolut MRR score.

In summary, the main contribution of this paper is a conversational multi-query rewriting method, CMQR, that can be utilized in conversational passage retrieval and applied on top of any pipeline that uses generative QR.  The novelty of our approach is twofold: (1) it generates multiple query rewrites at no extra cost compared to current neural QR approaches, (2) it effectively utilizes the generated rewrites in both sparse and dense retrieval.  Using the QReCC dataset for evaluation, we show that applying our method on top of any pipeline featuring generative QR improves performance, resulting in state-of-the-art results.

All resources developed for this paper (source code, query rewrites, and rankings) can be found at \url{https://github.com/iai-group/sigir2024-multi-query-rewriting}.

\section{Related work}
\label{sec:related}

We focus on the task of \emph{conversational passage retrieval}, where the goal is to retrieve relevant passages to the user query from a large passage collection.
Unlike generative approaches, here, hallucinations are ensured not to occur as answers can only come from the collection.  
While there is some variety between retrieval pipeline architectures for conversational search, the vast majority include QR, followed by a first-pass candidate selection stage and then by one or more re-ranking steps~\citep{Lajewska:2023:ECIR, Lin:2021:ACM,Yu:2020:SIGIR, Voskarides:2020:SIGIR, Kumar:2020:EMNLP}. This setup provides a good balance between efficiency and effectiveness~\citep{Asadi:2013:SIGIR}. 
We demonstrate the benefits of our approach to first-pass retrieval, using both spare and dense retrieval methods.

Automatic QR has a long tradition in IR, predominantly in query expansion, and has been shown effective in a range of tasks~\citep{Carpineto:2012:CS}.  Conversational query rewriting (CQR) aims to generate clear, de-contextualized queries from raw inputs by considering conversation context and addressing coreferences, ellipsis~\citep{Dalton:2019:TREC}, and topic transitions~\citep{Voskarides:2020:SIGIR}. Crucially, it helps clarify and refine the user's needs in a dialogue setting ~\citep{Penha:2019:arXiv}.
Recently, neural rewriting methods leveraging large pre-trained language models, like GPT-2~\citep{Kumar:2020:EMNLP, Vakulenko:2021:ECIR} and T5~\citep{Lin:2021:ACM}, have become prevalent.  While neural rewriting methods tend to outperform traditional query expansion techniques~\citep{Dalton:2020:TREC, Dalton:2021:TREC}, the best results are achieved by combining the two~\citep{Vakulenko:2021:ECIR, Lin:2021:ACM, Kumar:2020:EMNLP}.

Two previous CQR studies are particularly relevant to our work.  \citet{Lin:2021:ACM} propose two query reformulation methods: one focused on term importance and another on making human-like queries.  They show that fusing ranked lists after separate retrieval stages for both queries increases recall.  However, fusing the two lists after re-ranking showed no improvement.  The main difference between this work and ours is that we generate multiple natural language query rewrites.
\citet{Mo:2023:ACL} presents two neural models: one trained on rewriting queries and another to produce potential answers to the query, the idea being that pre-trained language models can directly answer questions by leveraging their internal knowledge. At inference, these potential answers are used to expand the query.  Our approach differs in that we use a single model to estimate term importance and pick expansion terms.

In another line of recent research, deep neural networks are used to generate query embeddings directly from context~\citep{Yu:2021:SIGIR, Mao:2022:SIGIR, Mao:2022:EMNLP}. These embeddings, used in conjunction with dense retrieval, can handle intricate conversational contexts more effectively.
While they integrate seamlessly with advanced neural models for IR, they require systems capable of interpreting them, potentially demanding more computational resources or additional processing steps. In contrast, the traditional approach of QR offers better interpretability and flexibility, translating complex conversational contexts into standalone, understandable queries that can be processed directly and efficiently with existing retrieval pipelines.

\section{Method}
\label{sec:method}

This section presents our method for generating multiple query rewrites in Section~\ref{sec:method:nqr}.  The integration of those rewrites in sparse and dense retrievals is described in Sections~\ref{sec:method:sparse} and \ref{sec:method:dense}, respectively.

\subsection{Conversational Multi-Query Rewriting}
\label{sec:method:nqr}

\subsubsection{Problem Statement}
\label{sec:method:nqr:ps}
\emph{Conversational query rewriting} (CQR) is the task of generating an informative context-independent query from a raw query (i.e., context-dependent user utterance) based on conversation context (i.e., history).
Formally, we let $q_i$ be the raw query at conversation turn $i$, and $H=\langle q_1, r_1, q_2, r_2, ..., q_{i-1}, r_{i-1} \rangle$ be the conversation history up to that point, where $r_j$ is a response provided by the system to the $j$th query ($j \in [1..i-1])$.
A context-independent query $\hat{q}_i$ is to created from the raw query $q_i$ by considering the conversation history up to that point: $\hat{q}_i = rewrite(\hat{q}_1, r_{1}, \hat{q}_{2}, r_{2}, \dots , \hat{q}_{i-1}, r_{i_1}, q_i)$.  The rewritten query $\hat{q}_i$ is considered self-contained and can be used downstream in various components of the retrieval pipeline.

\subsubsection{Motivation}
The majority of recent approaches employ generative neural models for CQR~\citep{Lajewska:2023:ECIR}.  However, these models often fail to find omitted information or detect topic shifts in longer conversations~\citep{Yan:2021:TREC}. In some cases, query rewrites introduce irrelevant terms, while in other cases, relevant terms are missing (akin to the notion of topic drift in pseudo relevance feedback~\citep{Macdonald:2007:CIKM,Shtok:2012:TOIS}).  
We hypothesize that it is often too challenging to accurately capture the user's information need in a single query rewrite.  Therefore, instead of returning a single most likely rewrite, we propose to return the top $n$ query rewrites generated with the same model and then utilize these rewrites in all stages of the retrieval pipeline.

More specifically, generative neural approaches to CQR commonly use the beam search algorithm~\citep{Graves:2012:arXiv}.  According to this technique, the probability scores of the $k$ most likely sequences are kept while generating a rewrite. When generation finishes, the sequence with the highest score is returned.

\subsubsection{Conversational Multi-Query Rewriting}

Motivated by the above, we propose \emph{conversational multi-query rewriting} (CMQR), which uses a fine-tuned generative language model to generate the top $n$ query rewrites at each turn $i$, $\hat{q}_i^1,\hat{q}_i^2,\dots,\hat{q}_i^n$, according to their beam search score. 
Each query rewrite $\hat{q}_i^j$ has an associated \emph{rewrite score}:
{\small
\begin{align*}
    RS(\hat{q}_i^j) = P(\hat{q}_i^j|H) = \left( \prod^{|\hat{q}_i^j|}_{l=1}P(t_l|t_{l-1}, \dots , t_{1}, H))\right)^{\frac{1}{|\hat{q}_i^j|}} ~,
\end{align*}
}%
where $t_1, \dots, t_l$ are the predicted tokens, $|\hat{q}_i^j|$ is the sequence length of the $j$th query rewrite, and $H$ is the conversation history (cf. Section~\ref{sec:method:nqr:ps}). Considering that RS is the product of the probabilities of all terms in a sequence, length normalization is applied to avoid the query rewriters' tendency to generate very short rewrites.

Due to the quadratic complexity with respect to the input size, a common practice is to limit the input to a maximum of 512 tokens~\citep{Lin:2020:arXiv}.  To accommodate this restriction, we limit the context to the previously rewritten utterances, $\langle \hat{q}_{1}, \dots, \hat{q}_{i-1} \rangle$, and the last system response, $r_{i-1}$.  We do not rewrite the very first user utterance of a conversation under the assumption that it is already self-contained and states the necessary context (i.e., $\hat{q}_{1} = q_1$). 

Next, we discuss how to utilize multiple query rewrites in various components of a retrieval pipeline.

\subsection{Sparse Retrieval}
\label{sec:method:sparse}

Sparse retrieval relies on a bag-of-words text representation, where each query term contributes to the document relevance estimate according to some scoring function, which is generally of the form
$score(q,d) = \sum_{t \in q} w_{t,q} \times w_{t,d}$, 
where $w_{t,q}$ and $w_{t,d}$ are the term weights associated with query $q$ and document $d$, respectively. 
Our interest is in setting the term query weights, $w_{t,q}$.  In the most commonly used retrieval scoring functions (e.g., BM25), this weight is taken to be the frequency of the term in the query, i.e., $w_{t,q} = c(t,q)$. 
In our approach, we construct a weighted bag-of-words query from all $n$ rewrites, where we set the weights for each term as the beam search score, i.e., $RS(\hat{q}_i^j)$. For each unique term in such obtained collection of terms, the term weights from all rewrites are summed up and normalized.

Effectively, the method performs both query expansion and a re-estimation of term importance based on multiple query rewrites.  A similarity can be drawn to relevance feedback algorithms like RM3~\citep{Lavrenko:2001:SIGIR}, where two weighted queries are interpolated: the original query and the relevance language model query. The difference is, here, we interpolate different queries extracted from conversational context instead of retrieved documents and do not assign a pre-determined portion of the total weight mass to the original query terms.
Importantly, our method is seen as complementary to relevance feedback and can be combined with it. %

\subsection{Dense Retrieval}
\label{sec:method:dense}

Dense retrieval differs from sparse retrieval in that it aims to compute a relevance score based on the similarity between queries and documents represented in a continuous embedding space instead of matching on exact terms.  In the simplest form, this score can be a dot product of the query and the document embedding vectors: $score(q,d) = h_q \cdot h_d$,
where $h_q$ and $h_d$ are the learned query and document embedding vectors, respectively.  We note that the learned embedding vectors can be pre-computed and stored for all documents in the collection, requiring only the computation of the query embedding vector at retrieval time.  

Given $n$ rewrites with associated weights, we first generate embeddings for all rewrites separately and scale them according to the associated weights. Then, the scaled embeddings are summed up into a single vector ($h_{q_i}$) that can be used in a regular dense retrieval system.  Formally, the query representation at turn $i$ is obtained by:
\begin{align*}
    h_{q_i} = \sum_{j=1}^n encode_q(\hat{q}_i^j) RS(\hat{q}_i^j) ~.
\end{align*}

\noindent
Essentially, the final query is a weighted centroid of the query rewrites.  This adds robustness to dense retrieval as the center of mass of multiple query rewrites will likely correspond better to the user's information need than a single rewrite would.

\section{Experimental Setup}
\label{sec:expsetup}

We present the datasets we use in our experimental evaluation, introduce our baselines, and provide implementation details.

\begin{table*}[ht]
\centering
\vspace*{\baselineskip}
\caption{Performance of sparse and dense retrieval with QR methods. Bold and underlined indicate the best and second-best results, respectively. $^*$ denotes significant improvements with a t-test at p < 0.05 of CMQR over its single-query counterpart.}
\label{tab:results}
\begin{tabular}{@{}cll*{12}{c}@{}}
\toprule
& \multirow{2}{*}{\textbf{Method}} & \multicolumn{3}{c}{\textbf{QReCC}} & \multicolumn{3}{c}{\textbf{QuAC}} & \multicolumn{3}{c}{\textbf{NQ}} & \multicolumn{3}{c}{\textbf{TREC-CAsT}} \\
\cmidrule(lr){3-5} \cmidrule(lr){6-8} \cmidrule(lr){9-11} \cmidrule(lr){12-14}
& & MRR & MAP & R@10 & MRR & MAP & R@10 & MRR & MAP & R@10 & MRR & MAP & R@10 \\
\cmidrule(lr){1-15}
\multirow{9}{*}{\rotatebox[origin=c]{90}{Sparse (BM25)}} & Manual rewrite & 39.81 & 38.45 & 62.65 & 40.32 & 38.98 & 62.90 & 40.78 & 39.05 & 63.80 & \textbf{27.34} & \textbf{27.04} & \textbf{53.77} \\
\cmidrule(lr){2-15}
& T5QR$_{Manual}$ & 31.03 & 29.86 & 50.17 & 30.75 & 29.60 & 49.77 & 34.06 & 32.51 & 52.79 & 24.15 & 23.91 & 46.90 \\
& ConQRR$^\dag$ & 38.30 & -- & 60.10 & 39.50 & -- & 61.60 & 37.80 & -- & 58.00 & 19.80 & -- & 43.50 \\
& ConvGQR & 49.18 & 47.66 & 68.01 & 51.34 & 49.82 & 70.10 & 45.57 & \underline{43.85} & \underline{64.06} & 25.91 & 25.20 & 47.26 \\
& LLM$_{adhoc}$$^\dag$ & 49.39 & 47.89 & 67.01 & \underline{53.01} & \underline{51.52} & 70.46 & 41.57 & 39.69 & 59.63 & 17.43 & 17.08 & 36.25 \\
& T5QR$_{LLM}$ & 46.72 & 45.19 & 64.00 & 50.13 & 48.64 & 67.50 & 39.26 & 37.14 & 55.99 & 16.98 & 16.97 & 34.64 \\
\cmidrule(lr){2-15}
& CMQR(T5QR$_{Manual}$) & 37.34$^*$ & 35.99$^*$ & 58.42$^*$ & 37.92$^*$ & 36.57$^*$ & 59.31$^*$ & 38.05$^*$ & 36.46$^*$ & 57.31$^*$ & 24.43 & 24.07 & 47.44 \\
& CMQR(ConvGQR) & \underline{50.24}$^*$ & \underline{48.79}$^*$ & \textbf{69.87}$^*$ & 52.41$^*$ & 50.98$^*$ & \underline{72.01}$^*$ & \textbf{46.83}$^*$ & \textbf{45.02}$^*$ & \textbf{65.34} & \underline{26.14} & \underline{25.78} & \underline{50.49} \\
& CMQR(T5QR$_{LLM}$) & \textbf{50.73}$^*$ & \textbf{49.2}$^*$ & \underline{69.25}$^*$ & \textbf{53.96}$^*$ & \textbf{52.50}$^*$ & \textbf{72.27}$^*$ & 44.11$^*$ & 41.95$^*$ & 62.40$^*$ & 20.78$^*$ & 20.46$^*$ & 43.80$^*$ \\
\midrule
\midrule
\multirow{9}{*}{\rotatebox[origin=c]{90}{Dense (GTR)}} & Manual rewrite & 43.15 & 41.27 & 66.12 & 40.67 & 38.92 & 64.59 & \textbf{54.01} & \textbf{51.25} & \textbf{73.13} & \textbf{43.74} & \textbf{42.98} & \textbf{65.23} \\
\cmidrule(lr){2-15}
& T5QR$_{Manual}$ & 36.08 & 34.41 & 56.95 & 33.70 & 32.16 & 55.36 & 46.11 & 43.65 & 63.50 & 38.11 & 37.30 & 58.76 \\
& ConQRR$^\dag$ & 41.80  & --  & 65.10 &  41.60  & --  & 65.90 &  45.30 &  -- &  64.10 &  32.70  & --  & 55.20 \\
& ConvGQR & 42.18 & 40.43 & 63.39 & 41.21 & 39.55 & 63.04 & 49.20 & 46.80 & 67.63 & 31.51 & 30.88 & 52.96 \\
& LLM$_{adhoc}$$^\dag$ & 44.99 & 43.19 & 67.34 & 45.21 & 43.48 & 68.30 & 47.64 & 45.20 & 67.27 & 30.91 & 30.48 & 51.03 \\
& T5QR$_{LLM}$ & 42.46 & 40.67 & 64.47 & 42.78 & 41.06 & 65.61 & 44.78 & 42.29 & 63.48 & 28.02 & 27.55 & 48.65 \\
\cmidrule(lr){2-15}
& CMQR(T5QR$_{Manual}$) & 40.53$^*$ & 38.73$^*$ & 63.15$^*$ & 38.72$^*$ & 37.02$^*$ & 62.12$^*$ & 48.50$^*$ & 46.01$^*$ & 67.67$^*$ & \underline{40.68}$^*$ & \underline{39.91}$^*$ & \underline{63.21}$^*$ \\
& CMQR(ConvGQR) & \underline{45.82}$^*$ & \underline{43.96}$^*$ & \textbf{69.75}$^*$ & \underline{45.00}$^*$ & \underline{43.20}$^*$ & \underline{70.00}$^*$ & \underline{51.95}$^*$ & \underline{49.50}$^*$ & \underline{71.17}$^*$ & 36.11$^*$ & 35.50$^*$ & 59.97$^*$ \\
& CMQR(T5QR$_{LLM}$) & \textbf{45.98}$^*$ & \textbf{44.17}$^*$ & \underline{69.31}$^*$ & \textbf{45.82}$^*$ & \textbf{44.08}$^*$ & \textbf{70.02}$^*$ & 49.58$^*$ & 47.14$^*$ & 69.18$^*$ & 34.69$^*$ & 34.09$^*$ & 57.64$^*$ \\
\bottomrule
\end{tabular}
\end{table*}

\subsection{Dataset \& Evaluation Metrics}

Following previous work~\citep{Wu:2022:EMNLP, Mo:2023:ACL, Ye:2023:EMNLP}, we use the QReCC~\citep{Anantha:2021:NAACL} dataset, which contains 14k conversations with 80k question-answer pairs, split into training and test sets (63.5k and 16.4k, respectively). The dialogues are based on questions from QuAC~\citep{Choi:2018:EMNLP}, TREC CAsT 2019~\citep{Dalton:2019:TREC}, and Google Natural Questions (NQ)~\citep{Kwiatkowski:2019:TACL}, with TREC CAsT appearing only in the test set.  Following \citet{Ye:2023:EMNLP}, test instances lacking valid gold passage labels are excluded from our analysis. Consequently, our dataset comprises 8,209 test instances, distributed as 6,396 for QuAC, 1,442 for NQ, and 371 for TREC-CAsT.
For a comprehensive evaluation, we present experimental results not only on the overall dataset but also on each subset.

We use mean reciprocal rank (MRR), mean average precision (MAP), and Recall@10 (R@10) as our evaluation metrics. 

\subsection{Baselines}
\label{baselines}

We consider the following baselines for comparison:
(1) \textbf{Manual rewrite}: Manually rewritten queries provided by the dataset.
(2) \textbf{T5QR$_{Manual}$}~\citep{Lin:2020:arXiv}: A strong T5-based~\citep{Raffel:2020:JMLR} QR model.  
(3) \textbf{ConQRR}~\citep{Wu:2022:EMNLP}: A T5-based model, optimized for retrieval performance using reinforcement learning.
(4) \textbf{ConvGQR}~\citep{Mo:2023:ACL}: An approach employing two T5-based models: one creates a de-contextualized query rewrite, the other predicts an answer to the query. The outputs are merged into a single query used for retrieval.
(5) \textbf{LLM$_{adhoc}$}\citep{Ye:2023:EMNLP}: An LLM query rewrite followed by an LLM query editor in an ad-hoc retrieval pipeline. The authors use ChatGPT 3.5 as their LLM.
(6) \textbf{T5QR$_{LLM}$}\citep{Ye:2023:EMNLP}: A sample of 10k datapoint is taken from the training set and run through the same approach as (4). The outputs are used to train a smaller, distilled model.

For each variant of the retrieval pipeline, we use the same T5-based QR approach, with a beam width of $k=10$, but consider only the top rewrite.  Wherever possible, i.e., code/model is made publicly available, we reproduce results on our system using V100 GPUs. Otherwise, we report the numbers from the original papers (indicated by $^\dag$).

\subsection{Implementation Details}

For QR, we fine-tune a T5 model~\citep{Raffel:2020:JMLR} starting from a \emph{t5-base}\footnote{\url{https://huggingface.co/t5-base}} checkpoint.  We set the beam width to $k=10$ for both single-query and multi-query approaches, as this was found to produce high-quality rewrites in~\citep{Lin:2021:ACM}.

For sparse retrieval, following~\citep{Anantha:2021:NAACL}, we employ the \emph{Pyserini}~\citep{Lin:2021:SIGIR} toolkit and use BM25 for retreival with hyparameters $k1 = 0.82$ and $b = 0.68$.  We generate dense embeddings using a GTR~\citep{Ni:2022:EMNLP} model from a publicly available checkpoint.\footnote{\url{https://huggingface.co/sentence-transformers/gtr-t5-base}} The implementation is based on \emph{Faiss}~\citep{Johnson:2021:IEEE}. 

\section{Results}
\label{sec:results}

The evaluation results on the QReCC test set along with a breakdown of specific subsets are reported in Table~\ref{tab:results}. The table is split into two groups: Sparse (Top) and dense retrieval (Bottom), with the same query rewriting methods in each group.  When multiple queries are considered for a given method, it is indicated by CMQR(*); in all our experiments, we consider 10 query rewrites.

Our main findings are as follows.
First, the CMQR method consistently outperforms both its sparse and dense retrieval counterparts across all datasets. This trend highlights CMQR's effectiveness in improving retrieval metrics. We show significant improvements in the range from 1.06 to 6.31 in sparse retrieval and 3.52 to 4.45 in dense retrieval in terms of MRR (absolute percentage points). The biggest improvement is observed when adding CMQR to the weakest model (T5QR$_{Manual}$), suggesting that the term importance and query expansion methods have a major effect. This observation is further supported by the smallest gain observed with ConvGQR, which has integrated query expansion.
Second, for both the sparse and dense retrieval groups, our CMQR method consistently outperforms all other methods on the overall QReCC dataset, achieving the best and second-best results, thereby setting a new state of the art. The strongest performance of CMQR with T5QR${LLM}$, a single \emph{t5-base} model, is notable for its compact size compared to previous best-performing methods, specifically the dual-\emph{t5-base} model (ConvGQR) and the two-step LLM model (LLM${adhoc}$). \citet{Ye:2023:EMNLP} show LLM$_{adhoc}$ is ~6 times slower than T5QR$_{LLM}$ making it impractical in real-world conversational applications.
Finally, manual rewrites, representing human effort in query rewriting, show strong performance, especially on the test-only TREC-CAsT dataset. However, CMQR methods still surpass these human efforts on the overall QReCC dataset, underscoring the potential of automated systems to enhance retrieval tasks. Interestingly, adding CMQR to the model trained on manual rewrites (T5QR$_{Manual}$) almost reaches the performance of manual rewrites, while CMQR applied to T5QR$_{LLM}$ shows an absolute improvement of 10.91 MRR percentage points compared to the manual rewrite.

\section{Conclusion}
\label{sec:concl}

In this paper, we developed a method for generating multiple query rewrites for conversational search and explored how these can be incorporated into sparse and dense retrieval. This approach differs from the majority of previous work, where only a single query rewrite is used.  We showed how multiple queries can be efficiently integrated at virtually no extra cost for both sparse and dense retrieval. Furthermore, we demonstrated that our method can be applied on top of existing query rewriting methods that employ generative query rewriting, yielding consistent improvements across all methods and resulting in state-of-the-art performance. 

In future work, we plan to employ multi-query rewrites also in the re-ranking components of multi-stage retrieval pipelines and determine automatically the number of rewrites to consider.

\begin{acks}
    An unrestricted gift from Google partially supported this research.
\end{acks}

\bibliographystyle{ACM-Reference-Format}
\balance
\bibliography{sigir2024-cmqr.bib}

\end{document}